\def\plotone#1{\centering \leavevmode
\epsfxsize=\textwidth \epsfbox{#1}}

\documentstyle[11pt, aas2pp4,psfig]{article}    

\def\sqig{$\sim$}

\def\degrees{$^{\circ}$}

\def\source{RX\,J0812.4-3114}
\def\src{RX\,J0812.4-3114}

\begin{document}
\title{
The Orbital Period of the Be/Neutron Star Binary
\src}

\author{Robin H.D. Corbet\altaffilmark{1} \&
Andrew G. Peele\altaffilmark{2}}

\affil{Laboratory for High Energy Astrophysics, Code 662,\\
NASA/Goddard Space Flight Center, Greenbelt, MD 20771}
\altaffiltext{1}{Universities Space Research Association;
corbet@lheamail.gsfc.nasa.gov}
\altaffiltext{2}{University of Maryland; current address University of Melbourne} 

\begin{abstract}
We present the results of Rossi X-ray Timing Explorer
observations of the Be star X-ray binary
system \src. A light curve obtained with
the RXTE All-Sky Monitor shows that the source is
currently in an active state with outbursts occurring
at approximately 80 day
intervals. The source underwent a transition from an
inactive state to this regular outburst state early in 1998. An
observation of \src\ was obtained with the RXTE Proportional
Counter Array
close to the time of a predicted maximum in March 1999 and strong
pulsations were detected at a period of 31.88 seconds. This confirms
the result of an earlier PCA observation by Reig \& Roche which
was serendipitously also obtained near the predicted maximum flux of
the 80 day period and also near the start of the current active state.
We interpret the periodicity in the ASM light curve as indicating the orbital
period of \src\ with outbursts occurring around periastron passage.

\end{abstract}

\section{Introduction}

\src\ was discovered by Motch et al. (1997) during a 
search for high-mass X-ray binaries by cross-correlating
SIMBAD OB star catalogs with low Galactic latitude
sources from the ROSAT all-sky survey. This X-ray source thus has an identified optical
counterpart, the Be star LS 992, and so it was suspected that this
source belongs to the high mass X-ray binary sub-class of Be/neutron
star systems. These systems are extremely variable and typically
transient. Their transient nature arises because of both the variable
envelope around a Be star from which a compact object accretes, and
the eccentric orbit of the compact object as accretion may often occur
only close to the time of periastron passage due to the increased
density of matter close in to the star and the magneto-centrifugal
barrier to accretion that operates at low
accretion rates (e.g. Stella, White \& Rosner
1986).

It it thus often possible to find the orbital period of a Be/neutron
star binary by searching for periodic outbursts in long term X-ray
observations. Once a binary period has been found this then facilitates
other studies and can aid in the detection of the pulse period by making it possible to schedule a pointed
observation at the time of periastron passage when the source is more
likely to be bright.

We report here on an analysis of the Rossi X-ray Timing Explorer (RXTE)
All-Sky Monitor (ASM) light curve of \src\ which shows a transition to
an active state in which regular outbursts occur. This is confirmed by
an RXTE Proportional Counter Array (PCA) pointed observation made at
the time of a predicted outburst during which strong pulsed X-ray
emission was detected.  This source has also been observed earlier with
RXTE by Reig \& Roche (1999; hereafter R2) who reported the detection of
31.88s period pulsations.

\section{Observations and Analysis}

\subsection{All-Sky Monitor}

The All-Sky Monitor (Levine et al. 1996) on board RXTE consists of
three similar Scanning Shadow Cameras, sensitive to X-rays in an energy
band of approximately 2-12 keV, which perform sets of 90 second pointed
observations (``dwells") so as to cover \sqig80\% of the sky every
\sqig90 minutes. The analysis presented here makes use of daily
averaged light curves constructed from the flux measured in individual
dwells. The Crab produces approximately 75 counts/s in the ASM over the
entire energy range and ASM observations of blank field regions away
from the Galactic center suggest that background subtraction may yield
a systematic uncertainty of about 0.1 counts/s (Remillard \& Levine
1997). The RXTE All Sky Monitor provides light curves of several
hundred sources including \src\ and we are regularly 
examining the light curves of
all these sources to search for periodic modulation and other
variability.

The ASM light curve of \src\ is shown in Figure 1.
It was noted early in 1999 (Corbet 1999) that since early 1998 the intensity of \src\ had
apparently increased and was exhibiting weak outbursts at intervals
of approximately every 80 days.
Based on the apparent \sqig80 day period in the ASM light curve we
predicted that the next maximum in the X-ray flux would occur at around
1999 March 25 and so RXTE PCA observations were undertaken to more
accurately measure the flux, investigate
the spectrum, and search for pulsations at that time.

\subsection{Proportional Counter Array}

The Proportional Counter Array (PCA) is described in detail by Jahoda et al. (1996).  This detector
consists of five, nearly identical, Proportional Counter Units (PCUs)
sensitive to X-rays with energies between 2 - 60 keV with a total
effective area of \sqig6500 cm$^2$. The PCUs each have a multi-anode
xenon-filled volume, with a front propane volume which is primarily
used for background rejection.  For the entire PCA across the complete
energy band the Crab produces a count rate of 13,000 counts/s.  The PCA
spectral resolution at 6 keV is approximately 18\% and the collimator
gives a field of view of 1\degrees\ full width half maximum.

Our observation of \src\ was performed on 1999 March 25
between 07:52 to 13:44 with 4 of the 5 PCUs
operating and a total observing time of 14.5 ks was obtained.
Spectra and light curves were extracted with the standard
set of FTOOLS software using only the top layers
of the PCUs in order to achieve better signal to noise. The background level
was estimated using ``pcarsp'' with a model appropriate for
low count rate sources.

\section{Results}

\subsection{Orbital Modulation in the ASM}

In order to quantify periodic modulation in the ASM light curve, and so
investigate the \sqig 80 day modulation that is
directly visible from
the light curve,
we calculated a weighted Fourier Transform of the
light curve and the resulting power
spectrum in shown in Figure 2.
Power can be seen at the lowest frequencies which is attributable
to the source transition from a low state to the repeated
flaring state after approximately MJD 50800 (early 1998). Aside from
this, the three highest peaks in the power
spectrum are at approximately 81 days ($f$ \sqig 0.012), the first
harmonic of this, and a sub-harmonic, thus confirming the
periodic modulation.

From both Fourier and $\chi ^2$ folding analyses we find a most
likely period for the flare recurrences of \sqig 81.3 days. Determining
a reliable error on this period estimate suffers from the problems that
only a small number of flares (nine) has been detected so far and that
individual outbursts from a Be star system may vary in shape and
perhaps phase of maximum flux caused by variations in the
Be star's circumstellar envelope (see e.g. Bildsten et al. 1997,
Wilson et al. 1997, Reig \& Coe 1999).
We thus estimate an error of \sqig1 day
on the orbital period . The epoch of maximum flux is also
estimated to be \sqig MJD 51260.5 $\pm$ 2.

The ASM light curve folded on the 81.3 day period is shown in Figure 3
with the data before and after the transition
to the active state folded separately.
In the active state the phase of X-ray activity appears to be concentrated in
a relatively narrow range and the average flux at maximum is roughly 10
mCrab.

We note that observations of \src\ made by R2 were
performed at the start of this
source's active period in early 1998 (February 1 and 3).
Serendipitously, the time of their observations was also close to the peak
of the outburst and the phase of the R2 observation corresponds to
approximately 0.9. The time of the R2 observation is also marked
in Figure 1.

\subsection{PCA - Pulsations}

We searched for pulsations from \src\ by calculating a power spectrum
from the PCA light curve and these were very strongly detected at a period
near 31.9 seconds.  To refine our estimate of the period we performed a
pulse time arrival analysis. First a template was constructed
by folding the light curve on the period found from the power
spectrum. The light curve was then divided into four sections
each of approximately equal duration
with breaks at the ends of ``good time'' intervals and these
sections of the light curve were also folded
on this period. The phase shifts between the four folded profiles and
the template were then calculated by
cross-correlation. A linear fit was made to the phase shifts found
and this was used to redetermine the period. This entire process was
then repeated using the refined period found from the linear
fit.
From this analysis we find a barycenter corrected pulse period
of 31.8856 $\pm$ 0.0001 s. This is comparable to the period of
31.8851 $\pm$ 0.0004 s measured by R2

The PCA light curve folded on our pulse period is shown in Figure 3. This
shows a complex profile which includes both a sharp
dip after the main peak and also a sharp secondary peak 0.5 in phase
away from the main peak.

\subsection{PCA - Spectrum}

To fit the energy spectrum of \src\ we adopted the same spectral model as used by R2 of an absorbed
power-law with a high-energy cut-off.
This gave somewhat similar results for the location of
the cut-off energy at 6.8 $\pm$ 0.5 keV compared
to 4.9 $\pm$ 0.4 keV. However, a significantly higher absorption
was required of 3.7 $\pm$  0.9 x 10$^{22}$ N$_H$ compared to 
the R2 value of 0.5 $\pm$ 0.3. In addition, our spectrum
is somewhat steeper with a power-law index of 1.42 $\pm$  0.1
compared to 1.0 $\pm$ 0.1.
Our best-fit model
gives an unabsorbed 3 to 30 keV flux of 2.42 $\times$ 10$^{-10}$ ergs cm$^{-2}$ s$^{-1}$ which
corresponds to an unabsorbed luminosity of 2.3 $\times$ 10$^{36}$ (d/9$kpc$)$^2$ ergs s$^{-1}$; this is approximately twice the luminosity reported by R2.

\section{Discussion}

We find strong evidence for the presence of a \sqig 80 day
period in the ASM light curve of \src.
By comparison with other Be star X-ray binaries, the time of maximum
flux is likely to coincide with periastron passage of a neutron
star. 
The orbital period of \sqig80 days that we find combined with the \sqig32
second pulse period is consistent with the correlation between
orbital and pulse period that is found for the majority of Be/neutron
star binaries (cf. Corbet 1986). 

RXTE is measuring the orbital and pulse periods of a steadily growing
number of high mass X-ray binaries. It is hoped that as the number of
systems with measured parameters increases, including the identification
of optical counterparts, this
will help to elucidate why many systems like \src\ do
follow the general correlation trend in the period-period diagram while other
systems such as X0726-260 and GRO J2058+42 may not (Corbet \& Peele 1997,
Corbet, Peele, \& Remillard 1997). It remains to be
determined whether these two systems have unusually short orbital
periods compared to what would be predicted
from the length of their pulse periods or, instead can, at least at times display two outbursts per
orbit (e.g. Wilson et al. 1998).

The larger absorption that we measure from our PCA observation compared
to that found by R2 could be due to our observations being performed closer to
the time of periastron passage and thus when more material surrounds
the neutron star.

The pulse profile of \src\ is surprisingly complex and will
require detailed modeling. The sharp dip after pulse maximum
may be due to an occultation of the pole by the body of the
neutron star while the secondary pole could be due to emission
from the other pole.

At the time of writing, \src\ continues to exhibit periodic outbursts.
If these persist this will allow both a refinement of the
orbital period and provide further opportunities for
studies with pointed instruments. During this active period
an envelope should be present around the Be star primary and
this should reveal itself through the presence of Balmer
emission lines in the optical spectrum.

\acknowledgments
This paper made use of quick look data provided by the RXTE ASM team at
MIT and GSFC. We thank our colleagues in the RXTE team for useful
discussions on several aspects of RXTE data analysis.

\pagebreak
\noindent
{\large\bf Figure Captions}

\figcaption[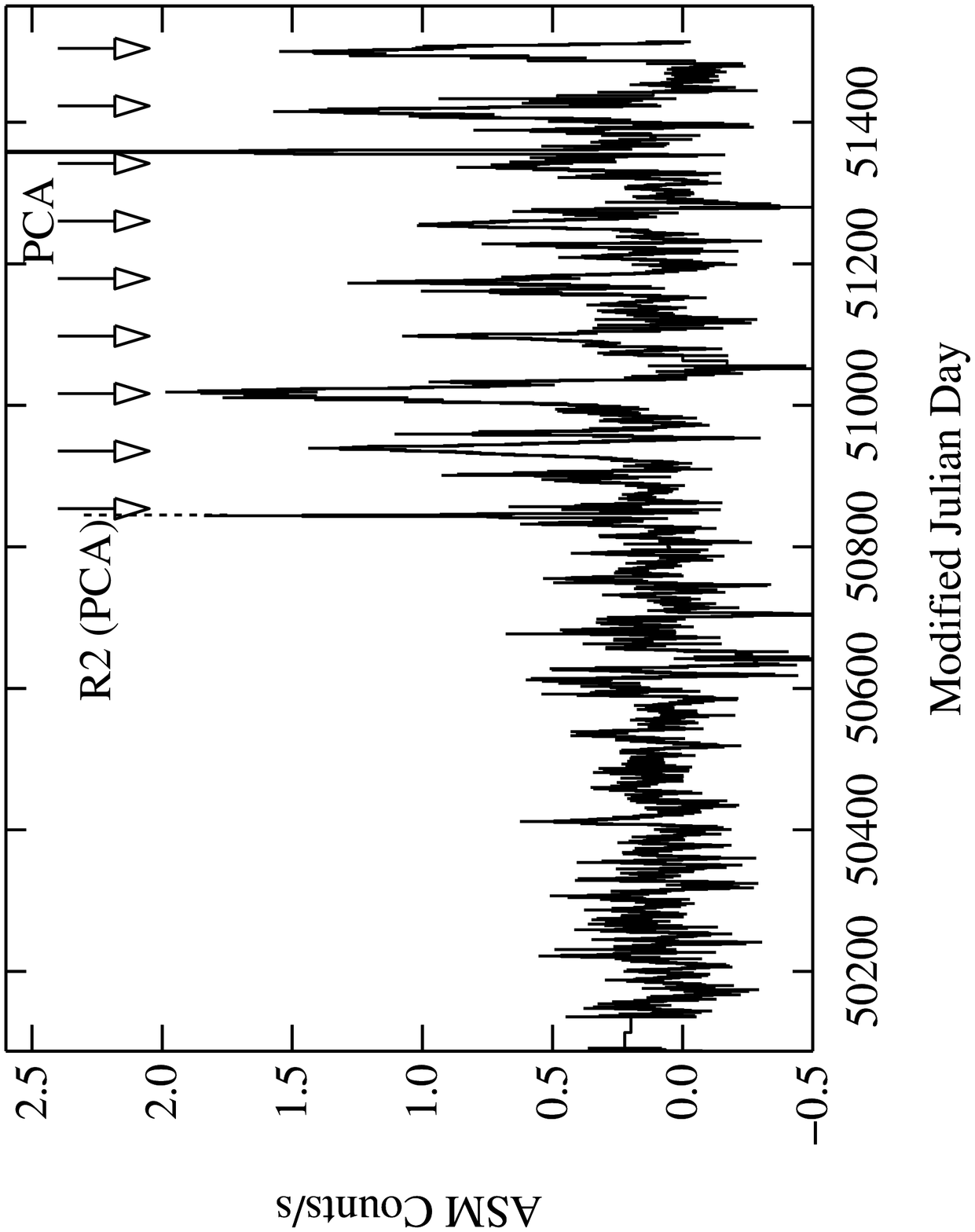]
{Light curve of \src\ obtained with the RXTE All Sky Monitor. This was
derived from the standard
daily averaged light curve by rebinning to two day bins and
then smoothing. The times
of expected outbursts since the start of 1998 are marked by the arrows
for an assumed period of 81.3 days and epoch of maximum flux = MJD
51260.5. The short dashed line shows the time of the first PCA
observation by Reig \& Roche (1999). The second PCA observation, which
is the one described in this letter, was obtained coincident with the
time of a predicted outburst.}

\figcaption[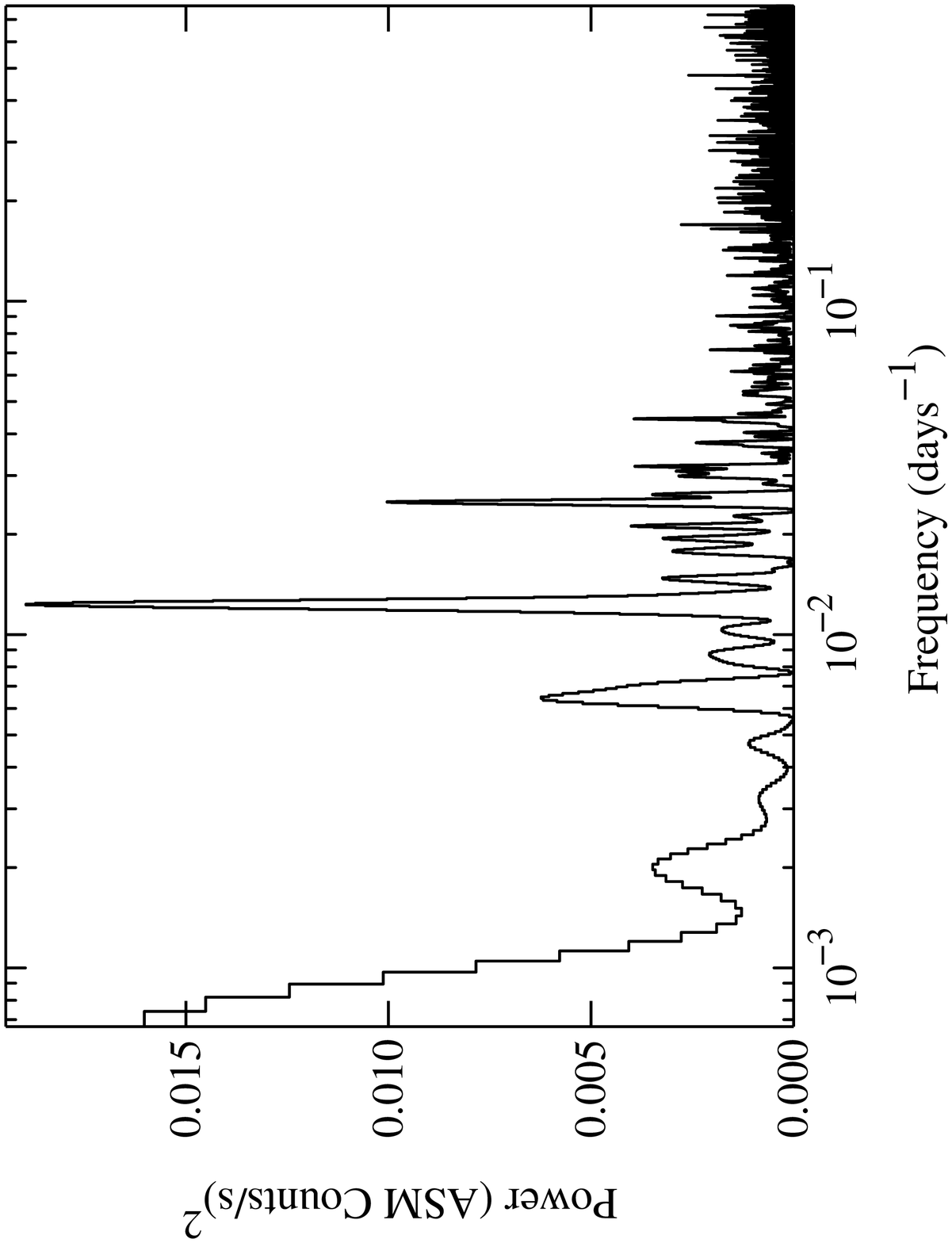]
{Power spectrum of the ASM light curve of \src. }

\figcaption[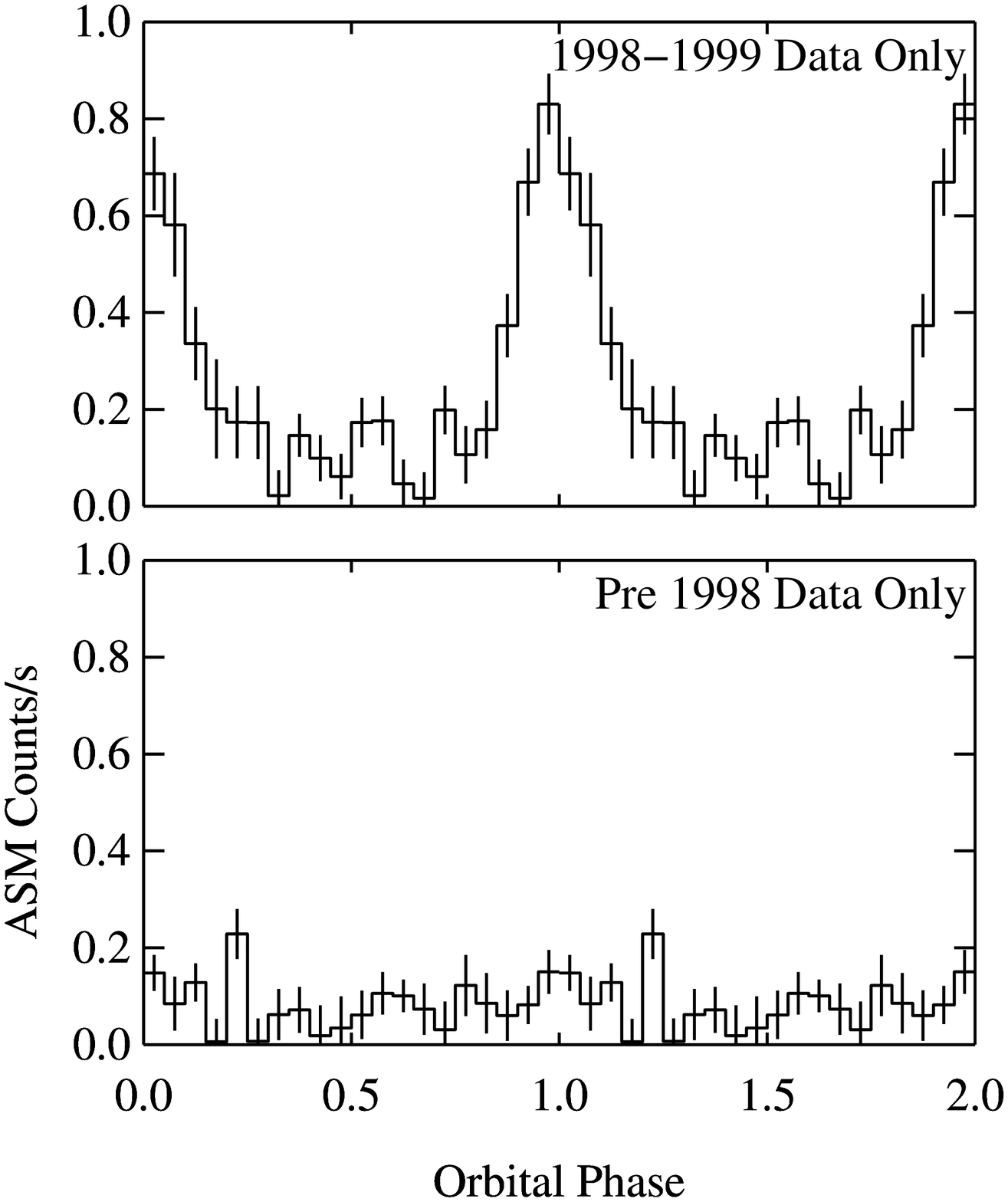]
{ASM light curve folded on the proposed 81.3 day orbital period. Folds
are done for before (lower panel) and after (upper panel) MJD 50800
illustrating the transition to an active state which occurred at around
that time.  Phase 0 corresponds to MJD 51260.5}

\figcaption[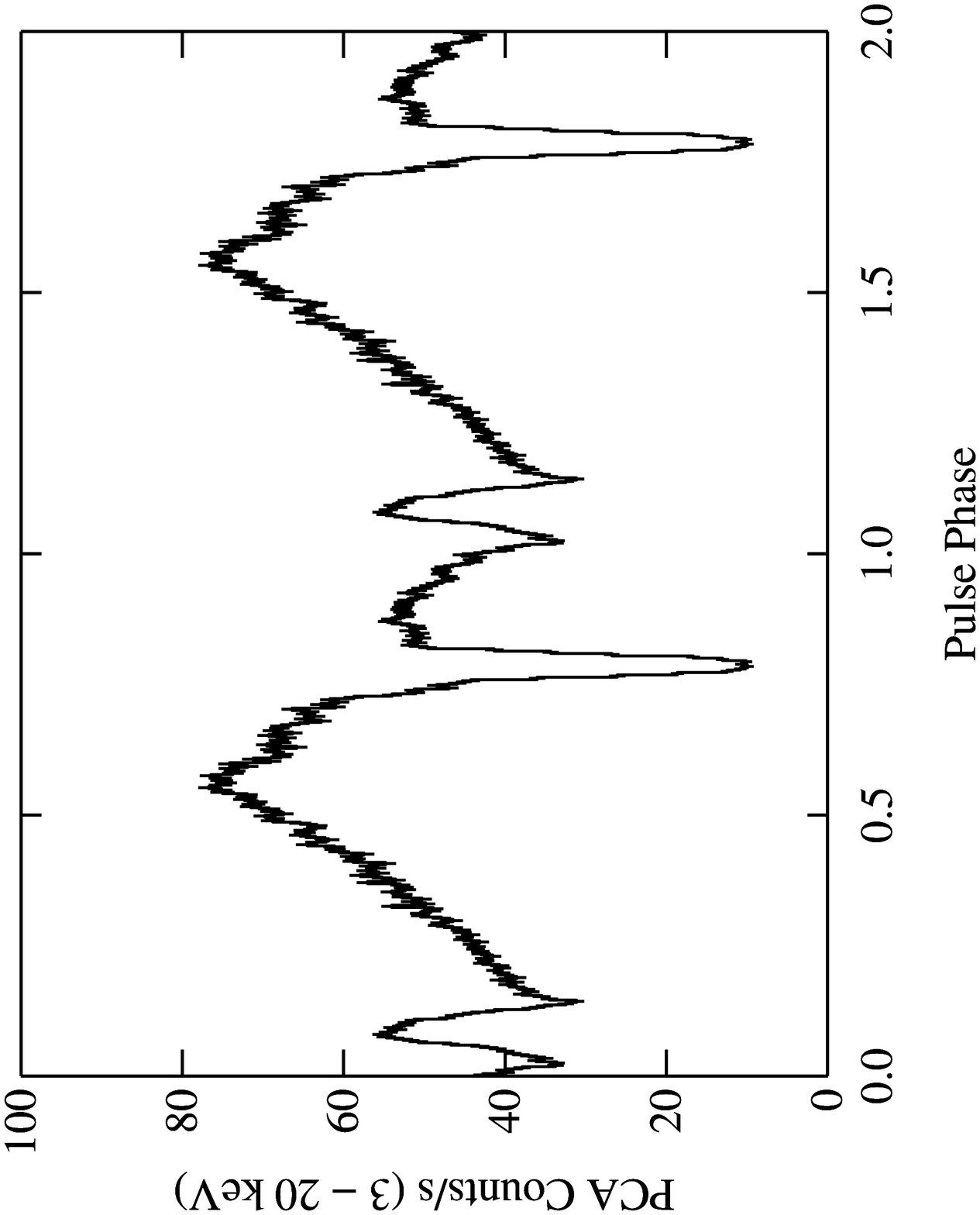]
{PCA light curve folded on the pulse period of 31.8856 s demonstrating
the complex pulse shape. The light curve was obtained from the four of
the five PCUs that were operational during this observation.
}

\begin{figure}
\plotone{figure1.ps}
\end{figure}


\begin{figure}
\plotone{figure2.ps}
\end{figure}


\begin{figure}
\plotone{figure3.ps}

\end{figure}


\begin{figure}
\plotone{figure4.ps}
\end{figure}

\end{document}